\documentclass[twocolumn,superscriptaddress,floatfix,preprintnumbers, nofootinbib,hyperref,groupedaddress]{revtex4-1} 
\pdfoutput=1
\usepackage[colorlinks=true,breaklinks=true]{hyperref}
\usepackage[normalem]{ulem}
\usepackage[utf8]{inputenc}
\hypersetup{allcolors=[rgb]{0.0 0.0 0.6},linkcolor=[rgb]{0.75 0.05 0.05}}
\usepackage{amsmath,amssymb}
\usepackage{epsfig}  
\usepackage{graphicx}   
\usepackage{slashed}       
\usepackage{tikz}
\usepackage{url}
\usepackage{color}
\usepackage{multirow}
\usepackage{comment}
\usepackage{amssymb}

\DeclareMathOperator{\cm}{cm}
\DeclareMathOperator{\GeV}{GeV}
\DeclareMathOperator{\eV}{eV}
\DeclareMathOperator{\meV}{meV}
\DeclareMathOperator{\keV}{keV}
\DeclareMathOperator{\MeV}{MeV}

\DeclareMathOperator{\s}{s}

\DeclareMathOperator{\kpc}{kpc}

\newcommand{\bp}{{\bf p}}

\newcommand{\beq}{\begin{equation}}
\newcommand{\eeq}{\end{equation}}

\allowdisplaybreaks

\bibpunct{[}{]}{,}{n}{}{,}

\begin{document}

\title{Detecting neutrino-boosted axion dark matter in the MeV gap}

\author{Pierluca Carenza}\email{pierluca.carenza@fysik.su.se }
%\affiliation[*]{The Oskar Klein Centre, Department of Physics, Stockholm University, Stockholm 106 91, Sweden}
\author{Pedro De la Torre Luque}\email{pedro.delatorreluque@fysik.su.se}
\affiliation{The Oskar Klein Centre, Department of Physics, Stockholm University, Stockholm 106 91, Sweden}
%ORCID: 0000-0002-4150-2539

%\date{\today}
\smallskip

\begin{abstract}
The elusive nature of Dark Matter (DM) remains a mystery far from being solved. A vast effort is dedicated to search for signatures of feeble DM interactions with Standard Model particles.
In this work, we explore the signatures of axion DM boosted by interactions with Supernova neutrinos: Neutrino-Boosted Axion DM ($\nu$BADM). We focus on $\nu$BADM converting into photons in the Galactic magnetic field, generating a peculiar gamma-ray flux. This signal falls in the poorly explored MeV energy range, that will be probed by next generation gamma-ray missions.
Once more, astrophysical searches might act as a probe of fundamental physics, unveiling the nature and properties of DM.
\end{abstract}

\maketitle

\section{Introduction}
Dark Matter (DM), precisely cold DM, is a key ingredient of contemporary physics to understand a series of astrophysical and cosmological observations~\cite{Simon:2019nxf,Salucci:2018hqu,Allen:2011zs,Bahcall:2000zu,White:1992ri,Hamilton:2000du,Arbey:2021gdg, Berezhiani:1989fp}.
However, its nature still remains an open problem. Remarkably, we do not have any indication that DM interacts with Standard Model (SM) particles~\cite{BERTONE2005279}.
Cosmic fluxes (as cosmic rays, gamma-rays and neutrinos) are a powerful tool to reveal whether DM is coupled with SM particles or not because, traveling on large distances, these particles maximize the probability of interactions with DM.

Many extensions of the SM include Axion-Like Particles, in the following simply `axions', light pseudoscalar particles with feeble interactions with ordinary matter.
For instance, effective theories derived from string theory predict several axions in a wide range of masses~\cite{Witten:1984dg,Svrcek:2006yi}, typically in the ultralight range, populating the so-called string axiverse~\cite{Arvanitaki:2009fg} (see also~\cite{Acharya:2010zx,Cicoli:2012sz}).
It is suggestive to explain the nature of DM with axions~\cite{Preskill:1982cy, Abbott:1982af, Dine:1982ah} (see also~\cite{Co:2020xlh,Arias:2012az,DiLuzio:2021gos,Blinov:2019rhb, Khlopov:1999tm} ). As a matter of fact, this DM candidate is receiving special attention in the last decade.
A common feature of many axions models is a coupling with SM fermions, important in astrophysical and laboratory searches. In particular, in this work we focus on the interesting and largely unconstrained interaction of axions with neutrinos~\cite{Reynoso:2022vrn,Reynoso:2016hjr,Huang:2018cwo,Guo:2022hnx}.

Here, we propose that neutrinos produced in a Galactic Supernova (SN) explosion might undergo elastic scattering with DM axions, boosting them and producing a relativistic flux of Neutrino-Boosted Axion DM ($\nu$BADM). We characterize the flux of $\nu$BADM for the first time, and show that the phenomenology associated to $\nu$BADM is broad. As a first preliminary study, we discuss the signatures of $\nu$BADM in presence of axion-photon interactions. Indeed, the vast majority of axion models predict an interaction with photons through the Lagrangian term
\begin{equation}
  \mathcal{L}_{\rm int} = \frac{1}{4} g_{a\gamma} a F_{\mu \nu} \tilde{F}^{\mu \nu} \,, \label{eq:lagph}
\end{equation}
where $F_{\mu\nu}$ is the electromagnetic tensor, $\tilde{F}_{\mu\nu}$ is its dual and $g_{a\gamma}$~denotes the axion-photon coupling. 
This coupling makes it possible for photons to oscillate into axions (and vice versa) in an external electromagnetic field. This peculiarity is at the foundation of experimental axion searches, e.~g.~\cite{Sikivie:1983ip,ParticleDataGroup:2022pth,Irastorza:2018dyq,ADMX:2003rdr,Lawson:2019brd}. For our purpose, we note that a fraction of the $\nu$BADM flux might convert into photons when traversing the Galactic magnetic field. The associated gamma-ray signal falls in the so-called MeV gap~\cite{Boddy:2015fsa}, in the reach of future gamma-ray experiments. 

This work is organized as follows. In Sec.~\ref{sec:BADM} we describe some important features of the $\nu$BADM flux. In Sec.~\ref{sec:detect}, we discuss the gamma-ray signal associated with $\nu$BADM converting into photons in the Galactic magnetic field. Moreover, we estimate the background to assess the detectability of such a signal in experiments at MeV energies. A description of these experiments is given in Sec.~\ref{sec:listexp}. In Sec.~\ref{sec:concl} we discuss the region of the axion parameter space accessible by these experiments and conclude.

\section{Supernova boosted axion dark matter}
\label{sec:BADM}

The recent idea that DM might be boosted by scattering with high-energy particles is applied in various cases, providing interesting phenomenology and promising detection perspectives, i.e.~\cite{Yin:2018yjn,Bringmann:2018cvk,Ema:2018bih,Cappiello:2019qsw,Dent:2019krz,Wang:2019jtk,Zhang:2020nis,Jho:2021rmn,Das:2021lcr,Chao:2021orr,Ghosh:2021vkt,Lin:2022dbl,PandaX-II:2021kai,CDEX:2022fig,Granelli:2022ysi,Cappiello:2022exa}.

In this work we discuss the possibility of boosting DM axions by means of elastic scattering with Supernova (SN) neutrinos. Indeed, SNe are famously known as neutrino factories since most of their energy is emitted in a neutrino burst of all flavors with $\mathcal{O}(10)\MeV$ energies~\cite{Janka:2006fh,Mirizzi:2015eza,Raffelt:1996wa}. The axion-neutrino interaction is dictated by the following Lagrangian~\cite{Huang:2018cwo}\footnote{Note that the choice of the derivative form of the Lagrangian in Eq.~\eqref{eq:axnu} is mandatory instead of the pseudoscalar form because in the considered process, $a\nu\to a\nu$, two axion lines are attached to a single fermion line~\cite{Raffelt:1996wa}. This conclusion is valid under the assumption that axions are Goldstone bosons.}
\begin{equation}
    \mathcal{L}=g_{a\nu}\bar{\nu}\gamma^{\mu}\gamma^{5}\nu\partial_{\mu}a\,,
    \label{eq:axnu}
\end{equation}
where $\nu$ is the neutrino field and $g_{a\nu}$ the dimensionful axion-neutrino coupling. We consider an axion model coupled only to tau neutrinos, a largely unconstrained interaction, and we consider $g_{a\nu}=3$~$\GeV^{-1}$, that is currently unconstrained but lies within the reach of future neutrino oscillation experiments~\cite{Huang:2018cwo}, like JUNO~\cite{JUNO:2015zny} and DUNE~\cite{DUNE:2015lol}. Precisely, in the mass range of interest the strongest constraint is the cosmological one, excluding $g_{a\nu}\gtrsim10\GeV^{-1}$~\cite{Huang:2018cwo}. 

The interaction in Eq.~\eqref{eq:axnu} allows neutrinos to boost axion DM through the Compton scattering $\nu a\to \nu a$. The spin-summed matrix element of the process, neglecting the axion mass, is calculated to be\footnote{The matrix element in Eq.~\eqref{eq:matel} does not take into account that axions interact with a neutrino flavor eigenstate, different from the mass eigenstate. To include this effect under our assumption of normal hierarchy, the matrix element in Eq.~\eqref{eq:matel} has to be multiplied by a factor $\cos^{2}\theta_{13}\cos^{2}\theta_{23}$ and consider only the heaviest mass eigenstate to contribute to the process. The relevant neutrino oscillations parameter are $\sin^{2}\theta_{13}=0.22$ and $\sin^{2}\theta_{23}=0.55$~\cite{ParticleDataGroup:2022pth}.}
\begin{equation}
\begin{split}
    \sum|\mathcal{M}|^{2}\simeq32g_{a\nu}^{4}m_{\nu}^{4}\frac{E_{a}^{2}}{E_{\nu}(E_{\nu}-E_{a})}\,,
\end{split}
\label{eq:matel}
\end{equation}
for a neutrino of mass $m_{\nu}$, with energy and momentum $E_{\nu}$ and $\bp_{\nu}$, respectively, interacting with an axion of mass $m_{a}$, initially at rest, and boosting it to an energy $E_{a}$. 
Since this interaction depends on the absolute value of the neutrino masses, we take these values to follow a normal hierarchy with values $m_{\nu,1}\simeq m_{\nu,2}\ll m_{\nu,3}=50\meV$, suggested by neutrino oscillation data~\cite{ParticleDataGroup:2022pth} and the absolute value of the heavy neutrino state is compatible with both laboratory~\cite{KATRIN:2019yun,KamLAND-Zen:2016pfg,GERDA:2020xhi} and cosmological constraints~\cite{Planck:2018vyg,eBOSS:2020yzd}.
The cross section of this process is easily obtained as
\begin{equation}
\begin{split}
     \frac{d\sigma}{dE_{a}}&=\frac{|\mathcal{M}|^{2}}{32\pi m_{a}|\bp_ {\nu}|E_{\nu}}\,,
\end{split}
\label{eq:cross}
\end{equation}
where the neutrino energy necessary to boost an axion to energy $E_{a}$ is $E_{\nu}\simeq m_{\nu}\sqrt{E_{a}/2m_{a}}$, making possible to transfer most of the neutrino energy to the axion for $m_{a}\gtrsim10^{-11}$~eV in the considered scenario.
We mention that a similar flux might be induced by different types of axion-neutrino interactions, as $\mathcal{L}\sim a^{2}\bar{\nu}\nu$ that could be interpreted as a low-energy effective interaction mediated by a heavy particle. For simplicity, in this work we only consider the more theoretically motivated interaction in Eq.~\eqref{eq:axnu}.

The energy-dependence of the resulting $\nu$BADM flux strongly depends on the properties of the incident SN neutrino flux. The time-integrated SN neutrino spectrum can be approximately described by
\begin{equation}
    \frac{dN_{\nu}}{dE_{\nu}}=C_{0}\left(\frac{E_{\nu}}{E_{0}}\right)^{\beta}e^{-(1+\beta)\frac{E_{\nu}}{E_{0}}}\,,
\end{equation}
where for the tau-neutrino spectrum $C_{0}=5\times10^{56}~\MeV^{-1}$, $E_{0}=15$~MeV and $\beta=1.3$~\cite{Fischer:2018kdt}. Note that we neglect the effect of neutrino oscillations on the SN neutrino spectrum since it would amount to a deviation smaller than $30\%$ from the considered spectrum and it is within the uncertainties of our calculation.
Here we assume that only $\nu_{\tau}$ interact with axion DM with the same coupling in Eq.~\eqref{eq:axnu}. Interactions with neutrinos of other flavors might lead to a comparable additional flux and, in the case of coupling with electron neutrinos, it might give an interesting phenomenology associated with the neutronization burst. We leave this possibility for future studies.

The cross section in Eq.~\eqref{eq:cross} is typically small, $\mathcal{O}(10^{-50}\cm^{2})$, but the huge neutrino flux\footnote{This is the reason why the Diffuse SN Neutrino Background~\cite{Lunardini:2005jf}, with a flux of $\lesssim 1~\cm^{-2}\s^{-1}$, is not expected to efficiently boost axion DM. Indeed, for comparison, the SN neutrino flux (for the parameters considered in this work) is $\sim1.6\times10^{13}~\cm^{-2}\s^{-1}$.} and the large number density of axion DM targets makes this interaction relevant in our context. Precisely, for an axion of mass $m_{a}\sim10^{-8}$~eV, the local DM number density would be $n_{a}=200\MeV\cm^{-3}/m_{a}\simeq\mathcal{O}\left(10^{16}\cm^{-3}\right)$. 

Within $1$~kpc from the Sun, there are 31 SN candidates~\cite{Mukhopadhyay:2020ubs}. The most famous is Betelgeuse, at $0.197$~kpc, and the closest one is Spica, at $0.08$~kpc. Considering a nearby Galactic SN ($d_{SN}\lesssim 1\kpc$) and approximating the DM density with the local one on these small scales, we calculate the time-integrated flux of $\nu$BADM per unit energy to be
\begin{equation}
    \frac{d\Phi_{a}}{dE_{a}}=\frac{d_{SN}n_{a}}{4\pi d_{SN}^{2}}\int_{E_{\rm min}}^{\infty} dE_{\nu}\frac{dN_{\nu}}{dE_{\nu}}\frac{d\sigma}{dE_{a}}\,,
    \label{eq:BADM}
\end{equation}
where
\begin{equation}
    E_{\rm  min}=\frac{E_{a}-m_{a}}{2}+\frac{1}{2}\sqrt{\frac{E_{a}^{2}m_{a}-m_{a}^{3}+2(E_{a}+m_{a})m_{\nu}^{2}}{m_{a}}}\,,
\end{equation}
is the lower limit of integration on the neutrino energy.
\begin{figure}[t!]
	\includegraphics[width=0.45\textwidth]{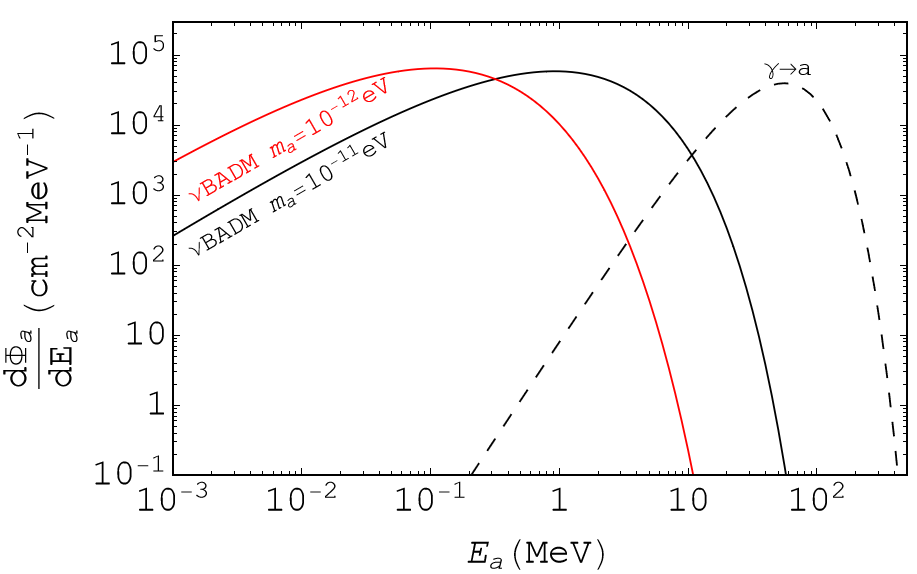}
	\caption{$\nu$BADM flux for the benchmark values $g_{a\nu}=3~\GeV^{-1}$, $m_{\nu}=50~\meV$ and two axion masses, $m_{a}=10^{-11}$~eV (black solid line) and $m_{a}=10^{-12}$~eV (red line). For comparison, the	SN Primakoff flux for massless axions with a coupling $g_{a\gamma}=10^{-12}\GeV^{-1}$ (black dashed line) is shown. These fluxes refer to a SN at $d_{SN}=0.197$~kpc.}
	\label{fig:comparison}
\end{figure}

The flux obtained from Eq.~\eqref{eq:BADM} is shown in Fig.~\ref{fig:comparison} for two representative axion masses, $m_{a}=10^{-11}\eV$ (black solid line) and $m_{a}=10^{-12}\eV$ (red line), and compared with the Primakoff flux of light SN axions with $g_{a\gamma}=$~$10^{-12}$~GeV$^{-1}$ (black dashed line). The considered SN is at a distance $d_{SN}=0.197$~kpc, as Betelgeuse. The energy range in which the $\nu$BADM flux falls depends on the energy of the incident neutrinos. Therefore, this flux is characterized by lower peak energies ($\mathcal{O}(\MeV)$) compared to an axion flux thermally produced in the SN core. 

\begin{table*}[t]
    \centering
    \begin{tabular}{|c|c|c|c|c|c|c|c|c|c|c|}
    \hline 
    Energy (MeV) & 0.05 & 0.1 & 0.5 & 1 & 5 & 10 & 50 & 100 & 500 & 1000 \\
    \hline
    % $\Phi_{Spica}$ (cm$^{-2}$ s$^{-1}$ MeV$^{-1}$) & 1.2e-1 & 3.86e-2 & 1.96e-3 & 5.72e-4 & 3.08e-5 &  8.02e-6 & 3.53e-7 & 1.06e-7 & 6.31e-9 & 1.50e-09 \\
    % \hline
    % $\mathcal{F}_{Spica}$ (events cm$^{-2}$) & 2.7e-3 & 1.41e-3 & 3.58e-4 & 2.10e-4 & 5.63e-5 & 2.94e-5 & 6.46e-6 & 3.88e-6 & 1.16e-6 & 5.49e-7  \\
    % \hline
    $\Phi_{Betelgeuse}$ (cm$^{-2}$ s$^{-1}$ MeV$^{-1}$) & 9.3e-2 & 2.98e-02 & 1.44e-03 & 4.16e-04 & 2.91e-05 & 1.0e-05 & 1.10e-06 & 5.04e-07 & 4.68e-08 & 1.14e-08 \\
    \hline
    $\mathcal{F}_{Betelgeuse}$ (events cm$^{-2}$) & 2.e-3 & 1.09e-03 & 2.64e-04 & 1.52e-04 & 5.33e-05 & 3.67e-05 & 2.02e-05 & 1.84e-05 & 8.57e-06 & 4.16e-06 \\
    \hline
  \end{tabular}
  
  \caption{Reference background gamma-ray flux and fluence estimated from the model developed by Ref.~\cite{delaTorreLuque:2022vhm} in the direction of Betelgeuse, for different energies around the MeV range. The fluence is estimated for $10$~s exposure time.}
    \label{tab:bkg}
\end{table*}

\section{Detectability of $\nu$BADM}
\label{sec:detect}

As discussed in the previous Section, in Fig.~\ref{fig:comparison} we notice that the $\nu$BADM flux is peaked in the few MeV energy range. This observation suggests that, if axions couple to photons through Eq.~\eqref{eq:lagph}, a fraction of the $\nu$BADM flux might be converted into photons while traveling in the Galactic magnetic field. The conversion probability in the direction of Betelgeuse, for example, is given by~\cite{Xiao:2020pra,Xiao:2022rxk}
\begin{equation}
\begin{split}
      P_{a\gamma}=&8.7\times10^{-6}\left(\frac{g_{a\gamma}}{10^{-11}\GeV^{-1}}\right)^{2}\left(\frac{B_{T}}{1{\rm\mu G}}\right)^{2}\\
      &\left(\frac{d_{SN}}{0.197\kpc}\right)^{2}\frac{\sin^{2}(qd_{SN})}{(qd_{SN})^{2}}\,,  
      \label{eq:convp}
\end{split}
\end{equation}
where $B_{T}=1.4~{\rm\mu G}$ is the modulus of the transverse Galactic magnetic field and the momentum transfer is
\begin{equation}
    qd_{SN}=\left[77\left(\frac{m_{a}}{0.1{\rm neV}}\right)^{2}-0.14\right]\frac{d_{SN}}{0.197\kpc}\frac{1~\keV}{E_{a}}\,.
\end{equation}
Eqs.~\eqref{eq:BADM}-\eqref{eq:convp} completely characterize the gamma-ray signature associated with the $\nu$BADM flux. We note that the expected gamma-ray emission from a core-collapse SNe is significant only after days from the bounce and at sub-MeV energies~\cite{Hjorth:2011zx, Alp_2019}.
Therefore, we expect a gamma-ray signal in coincidence with the neutrino burst to be originated only from exotic physics. 

We briefly comment that the Gamma Ray Spectrometer of the Solar Maximum Mission, that was, in principle, able to set an upper limit on the gamma-ray flux associated with $\nu$BADM in coincidence with SN 1987A~\cite{Brockway:1996yr,Grifols:1996id,Payez:2014xsa}. The upper bound on the gamma-ray flux in the energy range $4.1-6.4$~MeV sets a constraint on $\nu$BADM converting into photons. For this estimate we conservatively considered that neutrinos travel for $1$~kpc in the Large Magellanic Cloud, where the DM density is taken to be comparable with the local one ($\rho_{\rm DM}=200\MeV\cm^{-3}$~\cite{Read:2014qva}). %The obtained bound is not competitive with existing limits.
The obtained constraints lies below other existent astrophysical bounds~\cite{Reynolds:2019uqt,Dessert:2022yqq,Fermi-LAT:2016nkz,Dessert:2020lil,Payez:2014xsa,Caputo:2021rux}.

It is clear that to improve the detectability of this signal, experiments at MeV energies should be employed.
Indeed, the gamma-ray signal associated with $\nu$BADM conversion might be detectable in current or future experiments operative in this energy range, which is informally known as MeV gap~\cite{Boddy:2015fsa}, in view of the lack of observational measurements at these energies.  Indeed, since the Imaging Compton Telescope (COMPTEL)~\cite{COMPTEL} on board the Compton Gamma Ray Observatory (operating until the 2000), the gamma-ray sky above a few MeV remained mostly unexplored. 

To understand if the $\nu$BADM signal is observable, we have to estimate the number of background events that a detector would observe in coincidence with a SN explosion, during the $\sim10$~s window where we expect the neutrino burst and the associated $\nu$BADM flux. For this purpose we employ the model for the diffuse gamma-ray background derived in Ref.~\cite{delaTorreLuque:2022vhm} from the local flux of cosmic rays~\cite{Aguilar:2015ooa, aguilar2019towards} and the local HI gamma-ray emissivity spectrum~\cite{Casandjian:2015hja}. This model is able to reproduce the measured diffuse gamma-ray emission from the MeV~\cite{CGRO} within the uncertainties related to solar modulation~\cite{DelaTorreLuque:2022czl}. In addition, it also includes lines produced from the decay of unstable nuclei. We emphasize that the background emission below a hundred MeV is really uncertain due to the lack of experimental data and the uncertainties related to the modeling of that region. Therefore, this estimate is taken just to set a reference for the number of background events, but it should be remarked that uncertainties can be as large as a factor of $2$.

In Tab.~\ref{tab:bkg} we report the differential flux and fluence (number of events detected per unit of area) obtained from the reference background model. Namely, we report the average flux within $1^{\circ}$ (similar to the angular resolution of near-future detectors in the MeV region) around the position of the sky corresponding to  Betelgeuse ($b=9^{\circ}$ $l=199^{\circ}$), as a candidate for a nearby SN. Since the $\nu$BADM signal is expected to have a duration of around $10$~s (a burst for the detector), the fluence is computed for this exposure time. From Tab.~\ref{tab:bkg} we notice that the background decreases as the energy increases. Therefore, the detection strategy is to look at energies of a few MeV, to maximize the signal-to-noise ratio (SNR).

Once the background is characterized, in the next Section we summarize some running and future experiments able to probe this energy range.

\section{Experiments in the MeV gap}
\label{sec:listexp}
Nearly all current measurements of gamma-rays are limited to energies above hundreds of MeV, with most of the efforts devoted to identify signatures of Weakly Interacting Massive Particles (WIMPs) (see e.~g.~\cite{ParticleDataGroup:2022pth} and references therein). There has only recently been a renewed interest in exploring the few MeV range. Many proposed missions are designed to cover the MeV gap, 
% (e.g., MAST~\cite{MAST}, GRIPS~\cite{GRIPS} or GRAMS~\cite{GRAMS}, PANGU~\cite{PANGU}), AMEGO~\cite{AMEGO}, e-Astrogam~\cite{e-ASTROGAM}, AMEGO-X~\cite{AMEGO-X} or APT~\cite{Buckley2019Advanced}, among many others), what
showing the strong interest of the scientific community in observing this energy band (it is often said that MeV astronomy is for nuclear physics what optical astronomy is for atomic physics). As an example of the potential of observations in this energy range, we comment that different DM models could generally produce distinctive photon signatures at $\lesssim1$--$100$~MeV, such as lines or boxes~\cite{Mazziotta:2020foa}. 
% In fact, current direct-detection experiments are sensitive to DM masses above $1$~GeV, so their results do not apply to DM at much lower masses. 
Moreover, these measurements would improve our knowledge of the Galactic center (related to the Galactic Center Excess~\cite{Ackermann_2017}) and the injection of leptons from pulsars, among other relevant astrophysical phenomena.

In this section, we explore the detection perspectives of the proposed model for present and future gamma-ray missions, sensitive to the $\sim$~sub-MeV to GeV range. Tab.~\ref{tab:effarea} summarizes, for each experiment, the optimal energy range for the detection of $\nu$BADM and the corresponding averaged effective area. Moreover, we remark that the emission produced from axions thermally produced in the SN and converting into photons, which is peaked at higher energies and potentially detectable in {\it Fermi}-LAT~\cite{Meyer:2016wrm}, would not be detectable in the MeV range.

\subsection{NuSTAR}

The Nuclear Spectroscopic Telescope Array (NuSTAR)~\cite{Nustar} is a high-energy (its efficiency is peaked in the range $\sim2-79$~keV) X-ray telescope that was launched in 2012 and mainly consists of two co-aligned grazing incidence telescopes able to extend the sensitivity to higher energies as compared to previous missions such as Chandra and XMM with better temporal resolution. Although NuSTAR was designed for relatively long observations (1 day - weeks in duration), it can be used for the observation of very fast transient emissions due to its good temporal resolution ($<0.1$~ms). 
NuSTAR has a maximum effective area of $\gtrsim800$~cm$^2$ at $\sim10$~keV while it goes down to an average effective area of $\sim200$~cm$^2$ between $20-80$~keV.

\subsection{COSI}
The Compton Spectrometer and Imager (COSI)~\cite{COSI, COSI2}, is a gamma-ray telescope expected to be launched in 2025. A first version of COSI, was successfully operating  aboard NASA’s super pressure balloon in 2016 for a $46$-day flight. The main goals of this mission are the determination of the nature of Galactic positrons, the study of stellar evolution and nucleosynthesis in the Milky Way, and measurements of the polarization of gamma-ray bursts (GRBs) and compact objects. Moreover, this experiment recently gained attention in the context of exotic physics searches~\cite{Caputo:2022dkz}.
COSI is able to perform measurements of the diffuse and transient events in the $0.2$–$5$~MeV region, having a good background rejection and allowing for detection of GRBs or other gamma-ray flares over $>50\%$ of the sky. Its effective area is about $8$~cm$^2$ in the range $0.5-1.5$~MeV. 
In fact, this mission seems to be optimal for the detection of the $\nu$BADM emission (see also Ref.~\cite{Caputo:2022dkz} for detailed information on other prospects for DM searches from COSI).

%\url{https://cosi.ssl.berkeley.edu/}

\subsection{MeVCube}
The MeVCube experiment is a Compton telescope based on the CubeSat standard that has been proposed to probe the $100\keV-1\MeV$ energy range~\cite{Lucchetta:2022nrm} and is expected to be a powerful instrument for transient observations. The average effective area in this range is approximately $A=6\cm^{2}$ for normal incidence gamma-rays in its standard configuration (called $6U$, which stands for $6$ Cubesat units combined), but this can be increased by using the combination of more modules. Current studied extensions consist of $12$ and $16$~units combined. The $12U$ extension of MeVCube is expected to have an average effective area of $\sim15$~cm$^2$. Its maximum effective area is of $20$~cm$^2$ for the $12U$ extension and of $10$~cm$^2$ for the $6U$ configuration at $1$~MeV.

\subsection{e-ASTROGAM}
e-ASTROGAM~\cite{e-ASTROGAM} is a $\gamma$-ray telescope, whose proposal is in review by ESA, operating from about $150$~keV to $3$~GeV by combining the detection of Compton photons ($0.15$--$30$~MeV) and pair photons ($> 10$~MeV), similarly to AMEGO. 
This experiment expects to explore the MeV gap with an improvement of one-two orders of magnitude in sensitivity compared to the current state of the art (particularly, COMPTEL) and achieve a great improvement on source localisation accuracy (angular resolution $\gtrsim 1^{\circ}$ for $0.3$--$2$~MeV. It is designed to substantially improve the characterisation of the emission of GRBs, disentangling the high energy prompt
emission from the afterglow componen and measure the delay time with respect to the prompt keV-MeV component.
Its effective area is $120$--$560$~cm$^2$ ($300$~cm$^2$ at $1$~MeV) with an average effective area from $3$ to $10$~MeV is around $100$~cm$^2$. Above $10$~MeV (the pair-production domain of the satellite) the effective area increases to above $10^3$~cm$^2$ in the energy range between $50$ and $3\times10^3$~MeV.

\subsection{ComPair}
The Compton-Pair Production Space Telescope (ComPair)~\cite{ComPair, Shy2022:2210.02962v1} is a mission-concept proposed as a prototype of the AMEGO experiment to investigate the energy range from $200$~keV to more than $500$~MeV with high angular resolution and much better sensitivity than COMPTEL (around a factor of $20$--$50$ better). It also operates detecting both Compton-scattering events at lower energy and pair-production events at higher energies. Its effective area is $50$--$400$~cm$^2$ below $10$~MeV and $200$--$1200$~cm$^2$ above.

\subsection{AMEGO}
The All-sky Medium Energy Gamma-ray Observatory (AMEGO)~\cite{AMEGO} (and its upgraded version, AMEGO-X\cite{Caputo2022AMEGO}) is a mission expected to provide essential contributions to multimessenger astrophysics in the next years. This satellite would operate both as a Compton and pair-conversion telescope with great sensitivity between $\sim200$~keV and $>5$~GeV. This experiment is focused on the observation of astrophysical objects that produce gravitational waves and neutrinos in the extreme Universe through the study of extreme environments, such as kilonovae and supernovae, gamma-ray bursts and active galactic nuclei. 
AMEGO has an effective area that ranges from $500$ to $1000$~cm$^2$ across four decades of energy. For events of $10$~MeV to $\sim5$~GeV its effective area is $\sim500$~cm$^2$. For events of $1$--$10$~MeV its effective area is $\sim300$~cm$^2$. For events labelled as non-tracked (from $0.1$ to $100$~MeV) the effective area is around $3\times10^{3}$~cm$^2$.

\subsection{GECCO}
The Galactic Explorer with a Coded
Aperture Mask Compton Telescope (GECCO)~\cite{GECCO} is a telescope designed to conduct high-sensitivity measurements of the gamma-ray light in the energy range from $\sim0.2$~MeV to $10$~MeV and create intensity maps with high spectral and spatial resolution ($\sim1-2'$ angular resolution in the full energy range in mask mode). While most of the gamma-ray missions covering the MeV gap mainly focus on a Compton telescope, the GECCO mission combines a Compton telescope with the photoelectric regime with a coded mask.
It will be able to perform sensitive observations of the sky at MeV energies with unprecedented high spatial resolution that allows disentangling sources from diffuse emission (with special interest on the Galactic centre, the origin of the Fermi Bubbles and that of the $511$~keV line). It will also focus on the study of Galactic winds  and the role of low-energy cosmic rays in Galactic evolution and their sources, while being able to precisely localise transients.
Preliminary simulations of the performance of this telescope shows that its operation in the $0.1$--$10$~MeV energy range will have an energy resolution of $< 1\%$ and that its effective area will be of $\sim1200$~cm$^2$ between $0.3$--$3$~MeV.

\subsection{Advanced Particle astrophysics Telescope}
The Advanced Particle astrophysics Telescope (APT)~\cite{Buckley2019Advanced} is a space-based mission proposed to explore the gamma-ray sky with two main goals: confirm or rule out the thermal WIMP
dark matter paradigm and localize the prompt electromagnetic counterparts of gravity-wave/neutron-star mergers. This mission is expected to improve Fermi~\cite{Fermi-LAT} sensitivity at GeV energies by one order of magnitude, while providing sub-degree MeV transient localization over a large field of view. 
It is designed to operate from $300$~keV to $10$~MeV for the Compton detector and from $20$~MeV to $1$~TeV for the pair-conversion detector. Its maximum effective area for Compton events is expected to be $\sim10^4$~cm$^2$ (200 times larger than that of COMPTEL's effective area~\cite{COMPTEL}) and of $\sim7\times10^4$~cm$^2$ above $100$~MeV. Therefore, this mission is expected to significantly improve the study of transient emissions beyond AMEGO or e-ASTROGAM.

\begin{table}[t!]
    \centering
    \begin{tabular}{|c|c|c|}
    \hline
          & $\Delta E(\MeV)$&$A_{\rm eff}(\cm^{2})$ \\
        \hline
      NuSTAR & 0.05-0.1 &200\\
      COSI&0.5-1&8\\
      MeVCube (12U)&0.5-1&15\\
      e-ASTROGAM&1-5&100\\
      ComPair&1-5&250\\
      Amego&1-5&300\\
      GECCO&1-5&$1.2\times10^{3}$\\
      APT&1-5&$10^{4}$\\
      \hline
    \end{tabular}
    \caption{Optimal energy range, $\Delta E$, for the detection of $\nu$BADM for each experiment discussed in the text. The average effective area, $A_{\rm eff}$, in the chosen energy bin is also shown.}
    \label{tab:effarea}
\end{table}

\section{Discussion and conclusions}
\label{sec:concl}

\begin{figure}[t!]
	\centering
	\includegraphics[width=1\columnwidth]{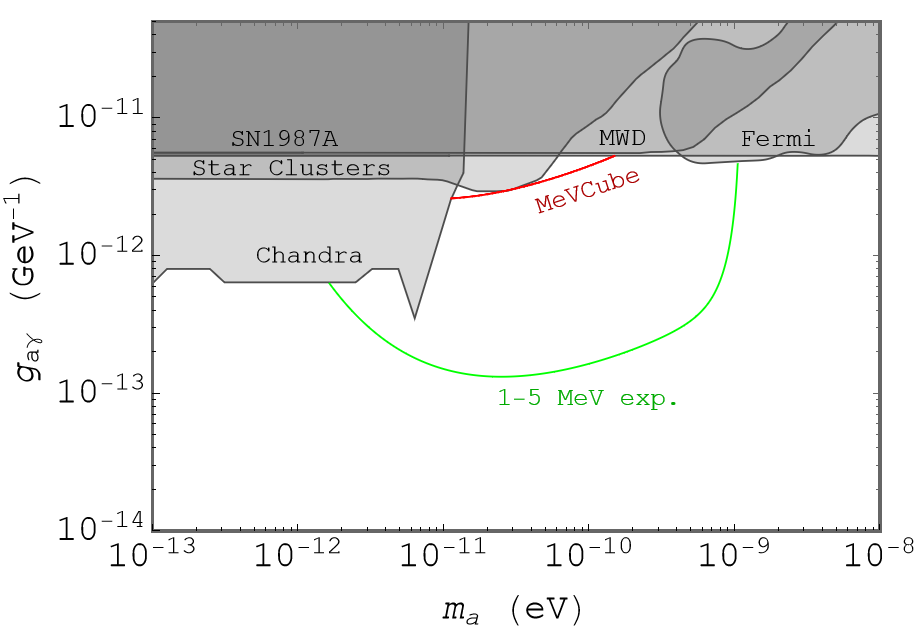}
	\caption{Region of the parameter space giving more than $1$ event in MeVCube 12U (red line) from $\nu$BADM converting into photons. For experiments in the $1-5$~MeV energy bin, we show the region with a SNR larger than $2$ (green line). 
	Here, we have considered a SN at $d_{SN}=0.197$~kpc, $g_{\nu a}=3$~GeV$^{-1}$ and $m_{\nu}=50$~meV. Astrophysical constraints listed in Refs.~\cite{Reynolds:2019uqt,Dessert:2022yqq,Fermi-LAT:2016nkz,Dessert:2020lil,Payez:2014xsa,Caputo:2021rux} are shown in grey.}
	\label{fig:bound}
\end{figure}

From the discussion above, we identify, for each experiment, the optimal energy range for $\nu$BADM detection. This is summarized in Tab.~\ref{tab:effarea}, with the corresponding effective area averaged over the energy bin.
Note that COSI and MeVCube have an effective area small enough that the expected background events are less than $1$, perhaps much less. In this case we assume that the detection of a single $\nu$BADM event is a sufficiently strong indication of non-standard physics. Therefore, in Fig.~\ref{fig:bound} we show a red line where we expect $1$ event observed in MeVCube, while the sensitivity region of COSI falls in an already excluded region of the parameter space.  

Experiments with a large effective area in the $1-5$~MeV range are ideal to probe $\nu$BADM. In these experiments, we expect more than $1$ event associated with the background. Therefore, with the green line we show where the SNR is equal to $2$, independently of the experiment. We emphasize, however, that the detection of the $\nu$BADM signal would be more significant for those detectors with higher acceptance area.
NuSTAR observations cannot set a constraint competitive with astrophysical limits at lower masses because of the limited effective area and large background.
Fig.~\ref{fig:bound} is obtained by considering a SN at $d_{SN}=0.197$~kpc, like Betelgeuse. 
The region potentially probed by a $\nu$BADM signal extends to low axion masses, down to \mbox{$m_{a}\sim 10^{-12}$~eV}, and small axion-photon coupling down to \mbox{$g_{a\gamma}\sim10^{-13}~\GeV^{-1}$}. 
We also highlight that the signature discussed in this work is different and complementary to the signal coming from axions thermally produced in the SN and converting into photons, which is peaked at higher energies and potentially detectable in {\it Fermi}-LAT~\cite{Meyer:2016wrm}. It might be relevant especially in the case of a detection of an axion-associated signal. In that case, by looking at lower energies it might be possible to understand another property of axions. 
Therefore it is important and timely to discuss the possibility of probing the nature of DM, the existence of axions and their properties thanks to the next Galactic SN explosion.

\section{Acknowledgements}
We warmly thank Tim Linden, Damiano F.~G. Fiorillo, Edoardo Vitagliano and Alessandro Mirizzi for useful comments on the manuscript.
The work of PC is supported by the European Research Council under Grant No.~742104 and by the Swedish Research Council (VR) under grants  2018-03641 and 2019-02337. PDLTL is supported by the Swedish Research Council under contract 2019-05135 and the European Research Council under grant 742104. This project used computing resources from the Swedish National Infrastructure for Computing (SNIC) under project Nos. 2021/3-42, 2021/6-326 and 2021-1-24 partially funded by the Swedish Research Council through grant no. 2018-05973

\bibliographystyle{bibi}
\bibliography{biblio.bib}

\end{document}